\documentclass[10pt, conference, compsocconf]{IEEEtran}

\let\labelindent\relax
\usepackage{enumitem}

\usepackage{amsmath}
\usepackage{graphicx}
\usepackage{multirow}
\usepackage{subfig}
\usepackage[linesnumbered, ruled]{algorithm2e}

\begin{document}

\title{\textit{Acceleration-as-a-Service}:\\Exploiting Virtualised GPUs for a Financial Application}

\author{
\IEEEauthorblockN{Blesson Varghese}
\IEEEauthorblockA{School of Computer Science\\
University of St Andrews, Fife, UK\\
Email: varghese@st-andrews.ac.uk}
\and
\IEEEauthorblockN{Javier Prades, Carlos Rea\~{n}o and Federico Silla}
\IEEEauthorblockA{Universitat Polit\`{e}cnica de Val\`{e}ncia\\46022 Valencia, Spain\\
Email: \{japraga,carregon\}@gap.upv.es, fsilla@disca.upv.es}
}

\maketitle

\begin{abstract}
\textit{‘How can GPU acceleration be obtained as a service in a cluster?’} This question has become increasingly significant due to the inefficiency of installing GPUs on all nodes of a cluster. The research reported in this paper is motivated to address the above question by employing rCUDA (remote CUDA), a framework that facilitates \textit{Acceleration-as-a-Service (AaaS)}, such that the nodes of a cluster can request the acceleration of a set of remote GPUs on demand. The rCUDA framework exploits virtualisation and ensures that multiple nodes can share the same GPU. In this paper we test the feasibility of the rCUDA framework on a real-world application employed in the financial risk industry that can benefit from AaaS in the production setting. The results confirm the feasibility of rCUDA and highlight that rCUDA achieves similar performance compared to CUDA, provides consistent results, and more importantly, allows for a single application to benefit from all the GPUs available in the cluster without loosing efficiency.
\end{abstract}


\IEEEpeerreviewmaketitle

\section{Introduction}
\label{introduction}
Hardware accelerators have found a prominent role in modern High-Performance Computing (HPC) solutions ranging from supercomputers to clusters. A number of supercomputers listed in Top500\footnote{http://top500.org} and Green500\footnote{http://www.green500.org} are supported by accelerators. For example, as of November 2014, two out of top six supercomputers listed on Top500 and the top ten supercomputers listed on Green500 have GPU accelerators. 

In HPC clusters, accelerators play an important role in allowing for heterogeneity of using both regular processors with accelerators \cite{hetcluster-1, hetcluster-2}. One way to set up such clusters would be to incorporate a GPU on each node of the cluster. While this set up can easily accelerate the computations of each node, it is not efficient because of relatively high performance/cost ratio of GPUs, higher power consumption of a node hosting GPUs \cite{gpu-power1}, and the under utilisation of GPUs when located on all nodes of the cluster.

An alternate and efficient cluster set up is to use fewer GPUs and provide nodes of the cluster access to the GPUs on-demand \cite{multitenancy-1, multitenancy-2}. The node can be treated as a client requesting \textit{Acceleration-as-a-Service (AaaS)} from a GPU server within the cluster. This results in one node receiving acceleration of multiple GPU servers as well as multiple nodes sharing the same GPU. Consequentially, the total power consumed in the cluster is lower than the former set up and increases the utilisation of GPUs in the cluster \cite{rcuda-energy}. 

Numerous challenges arise when developing multi-tenancy GPU frameworks, of which two are considered in this paper. Firstly, \textit{`how can GPUs on a server be made available as a service?'} Virtualisation technologies of GPUs can be used to bring this to fruition. The research we report in this paper explores the use of a framework that virtualises GPUs on a server and provides them as a service to nodes requesting them. The feasibility of the framework is tested on a real-world case study employed in the financial industry.

Secondly, \textit{`how can GPUs be efficiently shared between nodes?'} The cluster set up would not be efficient if virtualised GPUs could not be used concurrently and had to be exclusively locked for accelerating computations of a requesting node. In the research reported in this paper, acceleration can be requested as a service by any node in the cluster and concurrent access of a GPU by multiple cluster nodes is possible. 

The above challenges are addressed by the use of the rCUDA framework \cite{rcuda-new} which is reported in this paper. rCUDA facilitates placing requests for Acceleration-as-a-Service from  a client node without a GPU to a server node that actually hosts the physical GPU. The feasibility of the framework is tested on an application relevant to the financial risk industry. The application typically runs in a cluster environment, but can hugely benefit from GPU acceleration when it can be requested as a service for deriving important risk metrics in real-time. Experimental studies are performed on a cluster and the key result is that (i) rCUDA achieves near to similar performance as that of CUDA as there are very few overheads associated with using rCUDA over CUDA, (ii) consistent results can be obtained from rCUDA with negligible standard deviation, and (iii) multiple virtual GPUs can be used to provide acceleration to an application in order to boost performance. 

The remainder of this paper is organised as follows. Section \ref{framework} presents the rCUDA framework. Section \ref{casestudy} considers a financial risk application that is used to test the feasibility of rCUDA for providing Acceleration-as-a-Service in an HPC cluster. Section \ref{results} presents the platform and the experiments performed using the platform, and summarises the key results obtained. Section \ref{relatedwork} describes the related work in the area of HPC solutions for financial risk and virtualisation of GPUs. Section \ref{conclusions} concludes this paper by presenting future work.  

\section{The rCUDA Framework}
\label{framework}
The rCUDA framework, otherwise referred to as remote CUDA, is a middleware that facilitates the virtualisation of a CUDA\footnote{http://www.nvidia.com/object/cuda\_home\_new.html} (Compute Unified Device Architecture) compatible hardware accelerator, such as a GPU. The framework supports remote access to GPUs from a node of a cluster that does not have a physical hardware accelerator on it for accelerating computations of applications that run on it. In other words, acceleration is obtained as a service seamlessly to a requesting node without being aware of accessing remote GPUs. Furthermore, the source code of an application does not need any modification to reap the benefits of using rCUDA. The framework is freely available from http://rcuda.net/. 

The main benefits of rCUDA are: (i) Acceleration can be provided to an application as a service when requested. This is useful in enhancing on-demand computing seen in commodity clusters. A single application (running on a Virtual Machine (VM) or on a node of a cluster without a hardware accelerator) using rCUDA may benefit from the acceleration of a remotely located single GPU or multiple GPUs to reduce execution time, and (ii) increase the rate of GPU utilisation since multiple applications can access the same GPU. This in turn reduces the number of GPUs that need to be installed by an institution, and reduces the cost spent on energy consumption, cooling, physical space and maintenance, usually referred to as the Total Cost of Ownership (TCO).

In this section, we consider the architecture of rCUDA and the communication sequences in providing Acceleration-as-a-Service (AaaS). 

\subsection{Architecture}
\begin{figure}
\centering
	\includegraphics[width=0.45\textwidth]{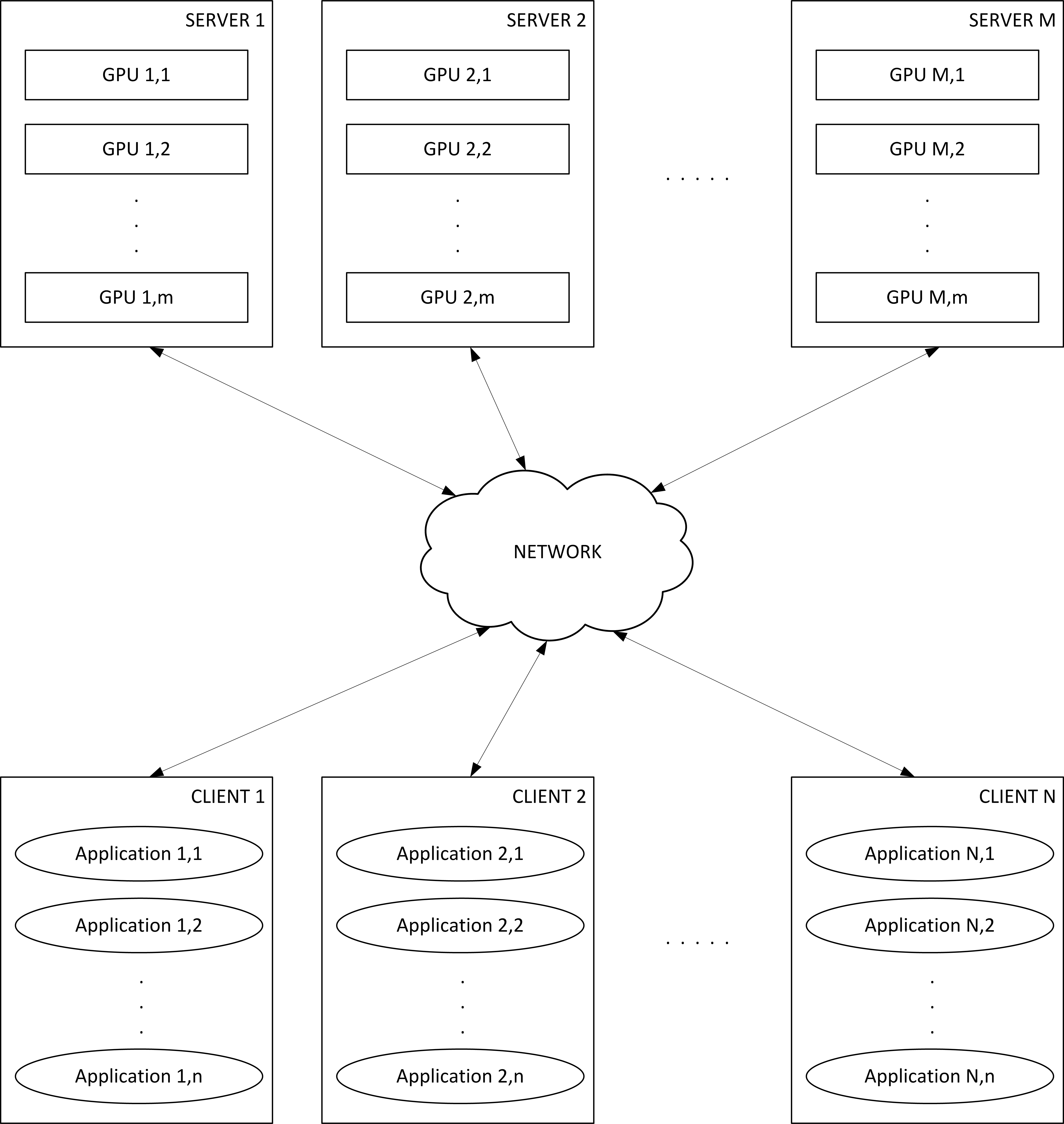}
	\caption{Distributed acceleration architecture facilitated by rCUDA}
	\label{figure1}
\end{figure}

The rCUDA framework aims to achieve a distributed acceleration architecture and is designed in keeping with the terminology familiar to the parallel and distributed computing communities. As shown in Figure \ref{figure1}, the rCUDA framework is a client-server architecture. Numerous \textit{Clients} executing applications that can benefit from hardware acceleration can concurrently access \textit{Servers} that have physical GPUs on them. The client makes use of the remote GPU to accelerate part of the software code of the application, referred to as kernel, running on it. The framework transparently handles the data management and the execution management; the transfer of data between the local memory of the client, the local memory of the server and the GPU memory, and the remote execution of the kernel.

\begin{figure}
\centering
	\includegraphics[width=0.48\textwidth]{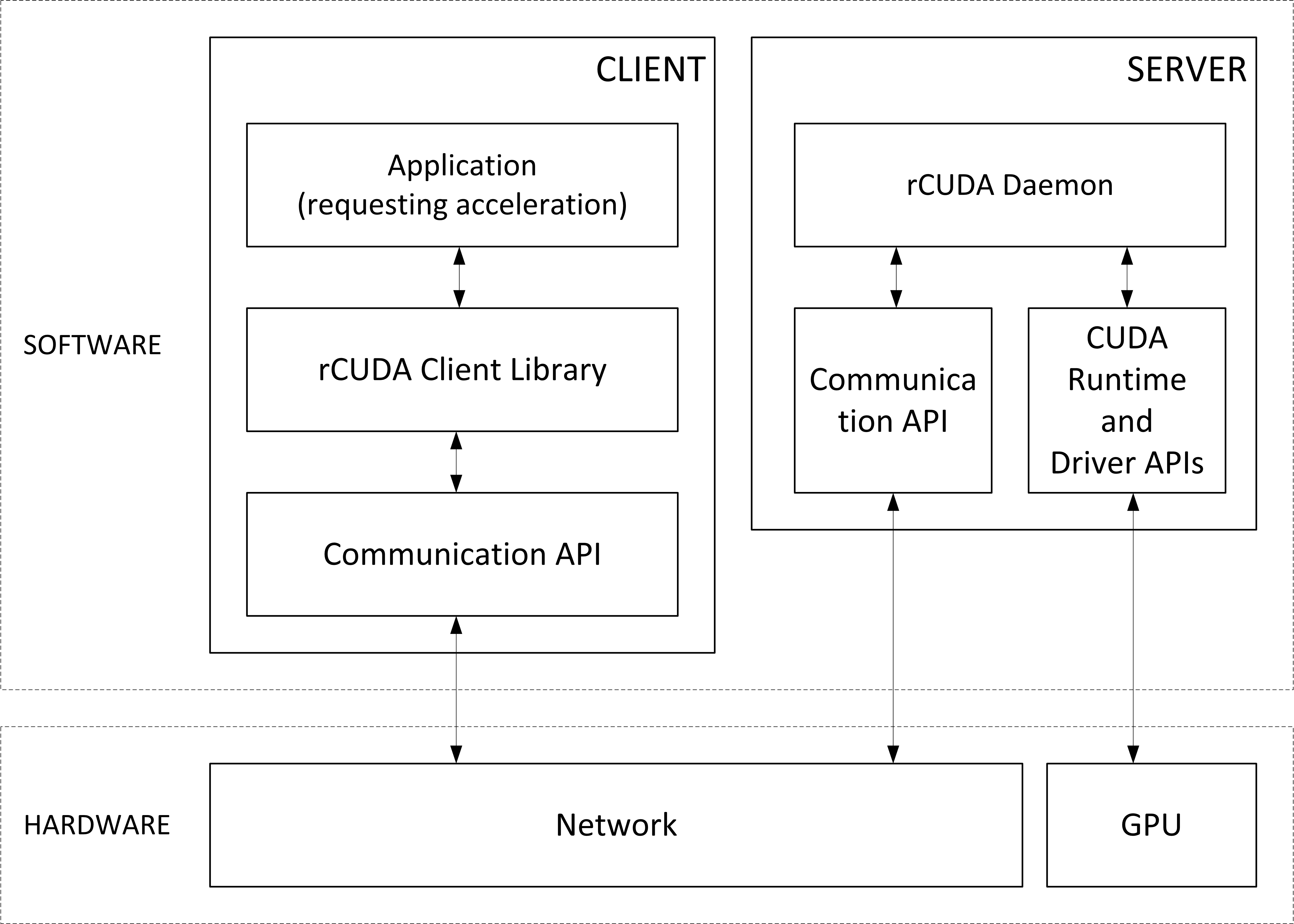}
	\caption{rCUDA client and server software/hardware stack}
	\label{figure2}
\end{figure}

\begin{figure}[t]
\centering
	\subfloat[An application with multiple threads accessing the same GPUs]{\label{figure2a}\includegraphics[width=0.48\textwidth]{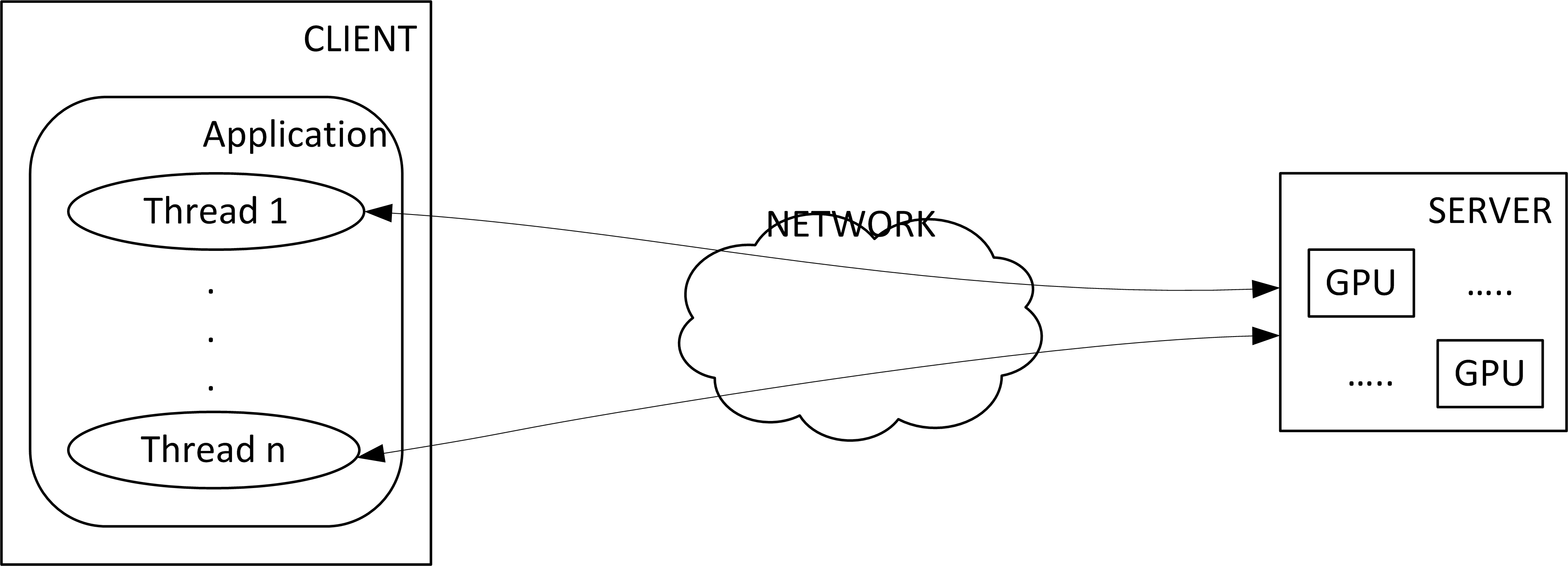}}\\
	\subfloat[An application with a single thread accessing multiple GPUs]{\label{figure2b}\includegraphics[width=0.48\textwidth]{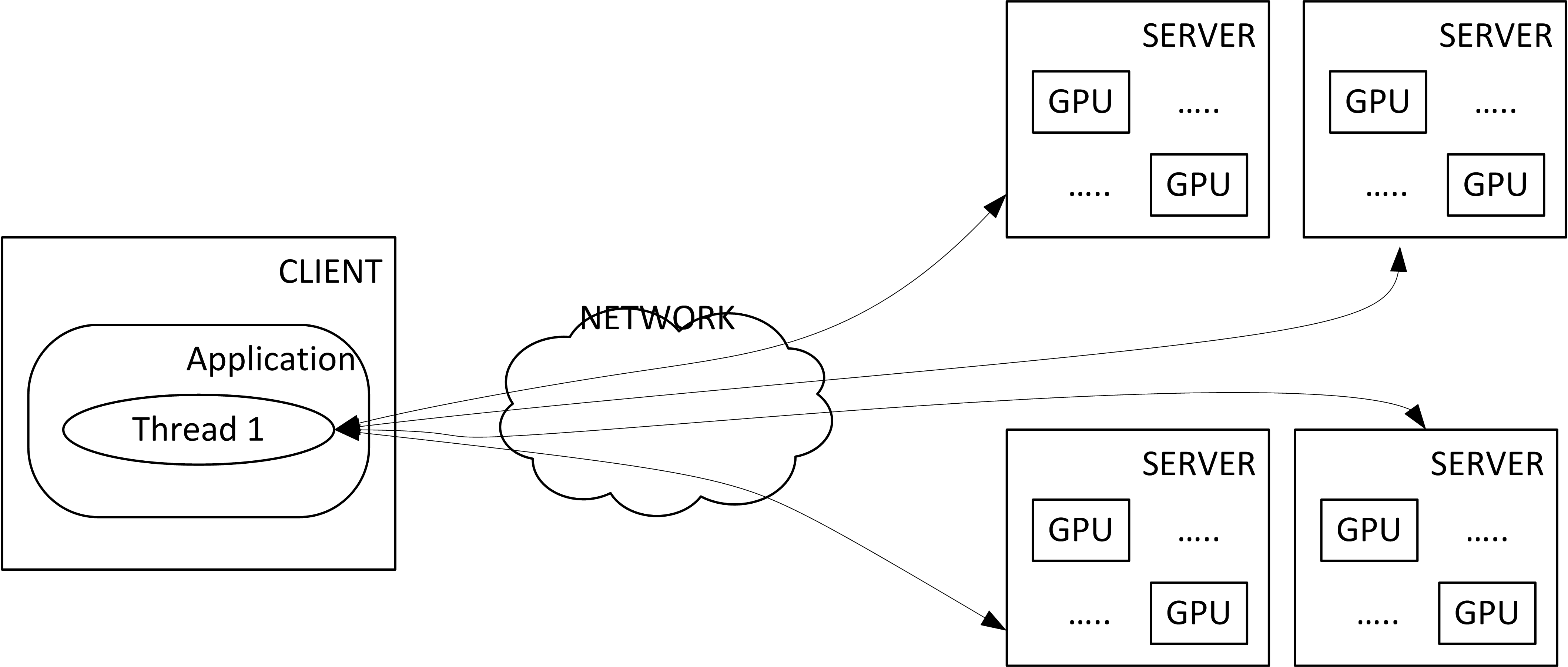}} \\
	\subfloat[An application with multiple threads each accessing multiple GPUs]{\label{figure2c}\includegraphics[width=0.48\textwidth]{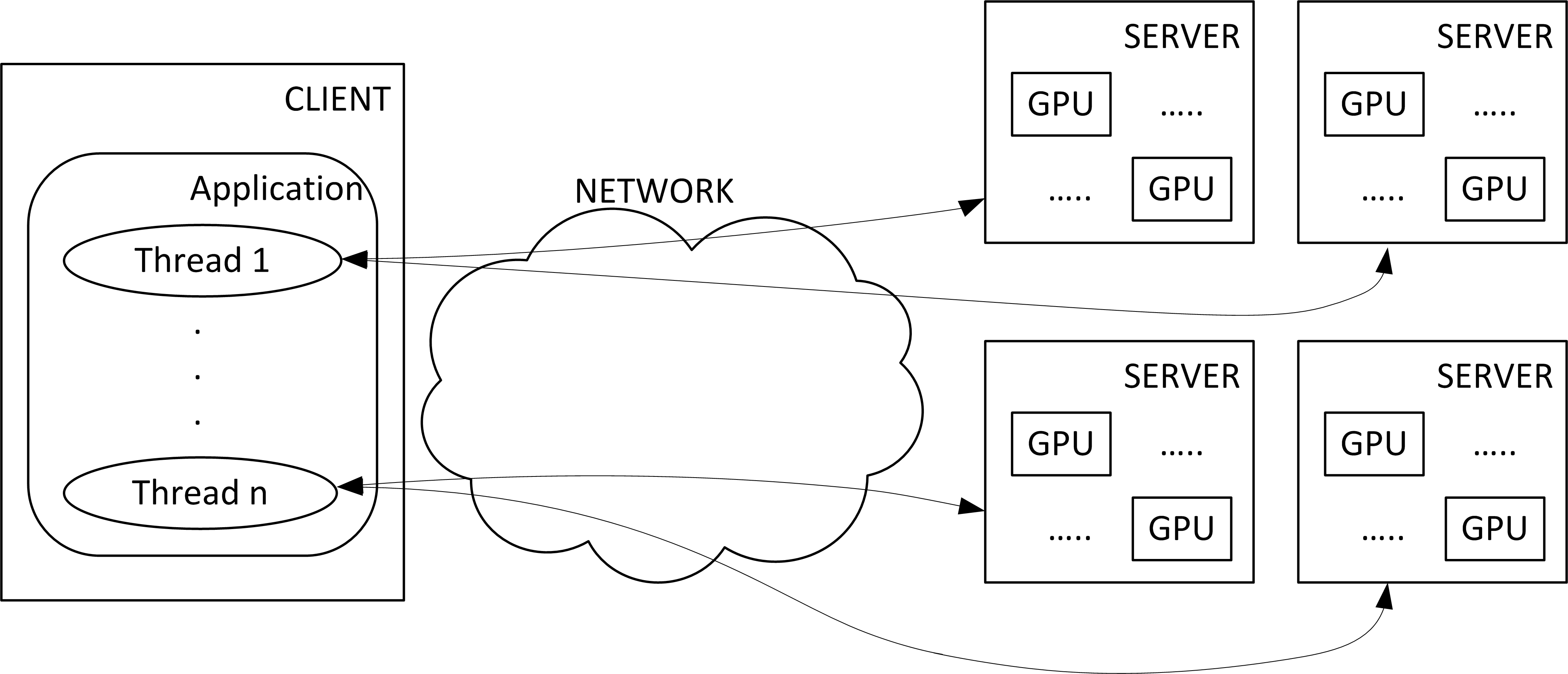}}
\caption{Different scenarios served by rCUDA}
\label{figure2A}
\end{figure}

Figure \ref{figure2} shows the hardware and software stack of the client and the rCUDA server. A proprietary API is used for communications across the network, using either regular TCP/IP sockets or the InfiniBand Verbs API when this high performance interconnect is available in the cluster. The rCUDA client and server will be considered later. 

rCUDA can serve different scenarios as shown in Figure~\ref{figure2A}. In the first scenario, a multi-threaded application on the client can request for acceleration, and if there is only one GPU in the cluster, then all threads are accelerated by the same GPU (Figure~\ref{figure2a}). It is noteworthy that a GPU is not exclusively locked to a thread and hence concurrent usage of the GPU is possible. In the second scenario, a single-threaded application on the client can request for acceleration, and if there are multiple servers hosting GPUs in the cluster, then the thread can leverage on using multiple GPUs (Figure~\ref{figure2b}). The third scenario (Figure~\ref{figure2c}) is a hybrid of the former scenarios, in which multiple threads of the same application are executed on single/multiple GPUs.

The rCUDA client and server shown in Figure~\ref{figure2} are now considered.
\subsubsection*{rCUDA Client}

The client nodes that execute the application (shown in Figure~\ref{figure1}), make use of the rCUDA Client Library, which is a wrapper around the CUDA Runtime and Driver APIs. The library is responsible for (i) intercepting calls made by the application to a CUDA device, (ii) processing them for forwarding the calls to the remote rCUDA server, and (iii) retrieving the results of the calls from the rCUDA server. 

Every client discovers the remote GPUs it can access in the cluster as if it were connected to its own PCIe port. When the wrapper is loaded on the client by the system dynamic linking loader, a connection to the server(s) is automatically established (which server(s) are specified in an environment variable). When the wrapper is unloaded the previously established connection to the server(s) are automatically closed and the resources which were required by the wrapper are released.  

For every CUDA call that is made by the application, the following tasks are performed by the rCUDA client to facilitate remote execution on the server. Firstly, function dependent local checks are performed, followed by mapping operations, such as assigning identifiers to pointers or to locally store information that is retrieved later on. A function identifier is used to pack the arguments of the function, and the execution request is then sent to the server. In the case of synchronous function calls, the client waits for a server response. 

\subsubsection*{rCUDA Server}
Each GPU server has an rCUDA daemon running on it which receives CUDA requests. The daemon then interprets the request and executes them on the physical GPU. For executing all requests from one client application on the GPU a new process is created on the server using a prefork technique as an independent GPU context. Hence, time-multiplexing on the GPU (or time sharing) is achieved by spawning a new server process for each remote execution over a new GPU context. In contrast to a multi-threaded solution, the rCUDA approach ensures that when a job is assigned to different GPUs, all co-processors can be safely shared by different jobs (provided there is sufficient device memory to run all applications concurrently). The proprietary device driver of NVIDIA manages the concurrent execution of active contexts using its own scheduler.

The prefork technique works as follows: (i) the parent server is started, (ii) a child server is created by a parent server which will serve all requests from a single remote execution, (iii) a connection request from a client is received by the child server, (iv) the parent server is notified when the connection is accepted by the child server, (v) the parent server then spawns another child server to serve subsequent requests from other clients, and (vi) child servers terminate after connection is closed by respective clients.  

\subsection{CUDA Management on rCUDA}
A CUDA program is similar to a C program executed on the CPU (host) and in addition contains functions executed on the GPU (device). The CUDA programming API provides functions similar to C as well as CUDA extensions. For example, CUDA functions such as \texttt{cudaMalloc}, \texttt{cudaMemcpy} and \texttt{cudaFree} are C like functions and the kernel launch using \texttt{\textit{kernel\_name}<<<blocks, threads>>>(\textit{parameters})} is a CUDA extension. 

A CUDA program is compiled using the NVIDIA \texttt{nvcc} compiler. Fragments of code that need to be executed on the device are compiled separately. During compilation a number of references to functions and structures are inserted into the host code. However, these functions and structures are not referenced in the CUDA documentation, which makes it almost impossible to create tools that will need to replace the NVIDIA CUDA Runtime Library. Solutions to mitigate this problem include the reimplementation of undocumented functions, for example, in GPU Ocelot \cite{gpuocelot}. However, these may not be compatible with future NVIDIA libraries since they are subject to change without notification. 

The rCUDA framework needs to manage the use of the aforementioned undocumented functions. The initial releases of rCUDA dealt with these undocumented functions in a very naive way: they were just not supported. This limited the use of rCUDA given that most CUDA applications use CUDA extensions to C. A compile-time workaround was created in later versions of the rCUDA framework \cite{rcudaconversion}, which transformed the CUDA extensions to C into regular C code, thus avoiding the use of the undocumented functions. However, although this approach was fully functional, the users were required to compile and execute the source code of applications for using rCUDA. The rCUDA framework was further developed to provide binary compatibility with CUDA applications; existing application binaries can be used with rCUDA without recompiling the source. 

\subsection{Acceleration-as-a-Service Communication Protocol}
An efficient communication protocol is developed for seamless execution between rCUDA clients and servers. 
This protocol is designed to provide lightweight support to the remote CUDA operations provided by the external accelerator. The CUDA commands intercepted by the rCUDA client wrapper are encapsulated into messages in the form of one or more packets that travel across the network towards the rCUDA server. The format of the messages depends on the specific CUDA command transported. In general, the messages have low network overheads. Every CUDA command forwarded to the remote GPU server is followed by a response message, which acknowledges the success/failure of the operation requested on the remote server. 

\subsection{Sample Communication Sequence}
In this section, the communication sequences between an rCUDA client application and an rCUDA server is illustrated. Matrix multiplication on a GPU is chosen as an example; consider two matrices, $A$ and $B$ that need to be multiplied and stored into matrix $C$. 

\begin{figure}
\centering
	\includegraphics[width=0.46\textwidth]{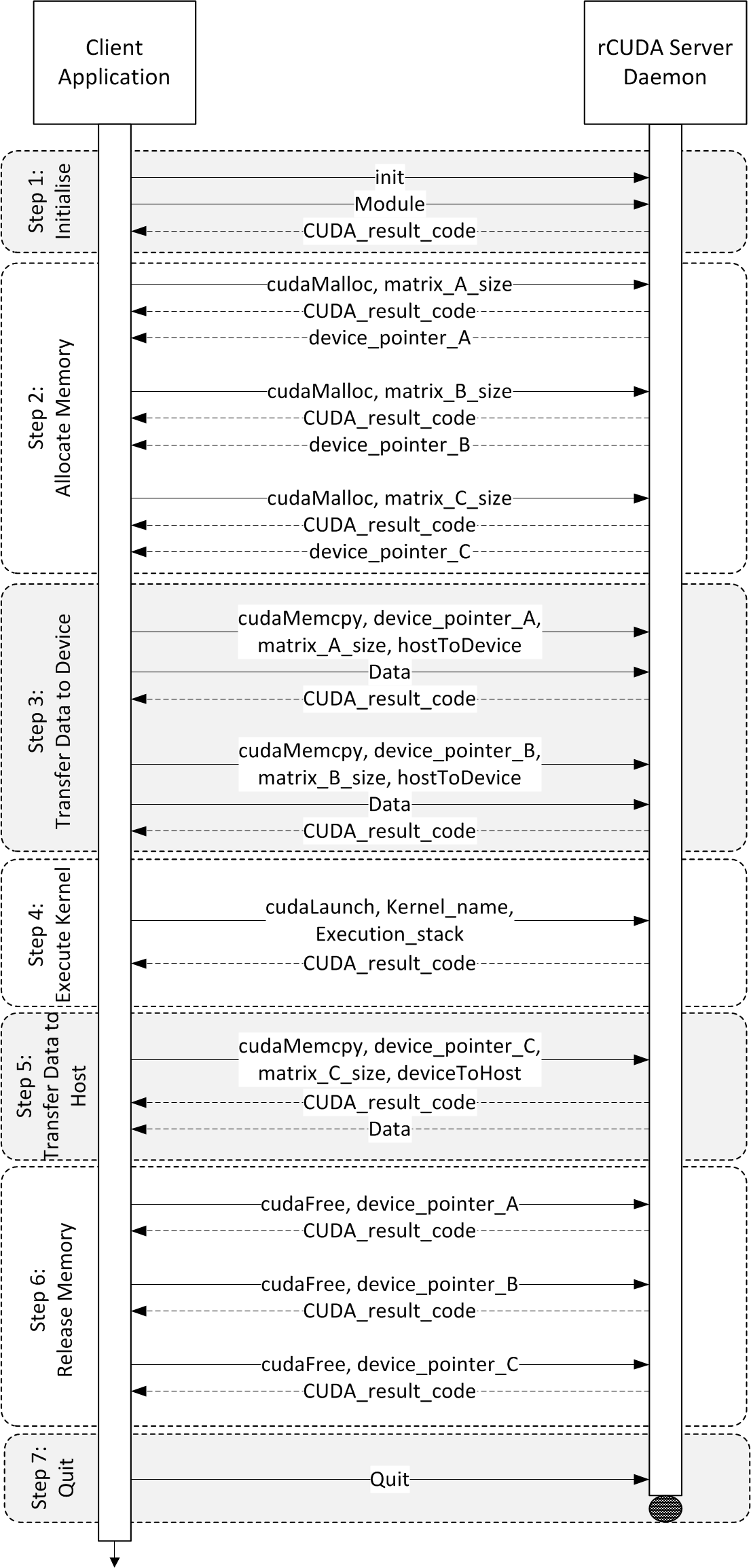}
	\caption{Sample communication sequence between an rCUDA client and server for matrix multiplication}
	\label{figure4}
\end{figure}

Figure \ref{figure4} shows the communication sequence between the client application and the rCUDA daemon executing on the remote server. The seven step protocol is as follows:
\subsubsection*{Step 1 - Initialise}
The client establishes connection with the remote server automatically, and the request for acceleration services is intercepted and the GPU kernel along with related information such as statically allocated variables are sent to the server.    

\subsubsection*{Step 2 - Allocate Memory}
Based on the client request memory is allocated on the GPU for data that will be required by the GPU kernel. Device memory is allocated for the three matrices, $A$, $B$ and $C$ on the remote server. Therefore, three \texttt{cudaMalloc} requests are intercepted by the client and forwarded to the remote server. 

\subsubsection*{Step 3 - Transfer Data to Device}
All data required by the kernel is transferred from the host to the remote device. 

\subsubsection*{Step 4 - Execute Kernel}
The GPU kernel is executed remotely on the rCUDA server. 

\subsubsection*{Step 5 - Transfer Data to Host}
After the execution of the kernel on the remote server the data is transmitted back to the host. 

\subsubsection*{Step 6 - Release Memory}
The memory allocated on the remote device is released. 

\subsubsection*{Step 7 - Quit}
In this final step the client application stops communicating with the remote server. The rCUDA daemon executing on the server stops servicing the execution and releases the resources associated with the execution.

\section{A Financial Risk Case Study}
\label{casestudy}
In this section, we demonstrate the feasibility of the rCUDA framework. For this we require a candidate application that can benefit from Acceleration-as-a-Service (AaaS) in HPC clusters. In this section, we present such an application employed in the financial risk industry, referred to as \textit{`Aggregate Risk Analysis'} \cite{s1} for validating the feasibility of rCUDA. The analysis of financial risk is underpinned by a simulation that is computationally intensive. Typically, this costly analysis is periodically performed on a routine basis on production clusters to derive important risk metrics. Such a set up is sufficient when the analysis does not need to be performed outside routine. However, risk metrics will need to be obtained in real-time settings, such as in online pricing, in addition to routine executions. In this context, the acceleration offered by virtual GPUs in a HPC cluster can be leveraged to develop a faster application fit for use in real-time settings. The rCUDA framework suits such an application because minimal changes need to be brought about to the production cluster and the acceleration required for the analysis is obtained as a service from a remote host. The analysis has previously been investigated in the context of many-core architectures \cite{s4}, but we believe virtual GPUs can be a better option. 

Aggregate risk analysis is performed on a portfolio of risk held by an insurer or reinsurer and provides actuaries and decision makers with millions of alternate views of catastrophic events, such as earthquakes, that can occur and the order in which they can occur in a year. To obtain millions of alternate views, millions of trials are simulated with each trial comprising a set of possible future earthquake events and the probable loss for each trial is estimated. 
Three data tables are required for the analysis, which are as follows:

\subsubsection*{Year Event Table} 
This is a database of pre-simulated occurrences of events from a catalogue of stochastic events that is denoted as $YET$. Each record in a YET called a `trial', denoted as $T_i$, represents a possible sequence of event occurrences for any given year. The sequence of events is defined by an ordered set of tuples containing the ID of an event and the time-stamp of its occurrence in that trial $T_i = \{(E_{i, 1}, t_{i, 1}), \dots, (E_{i, k}, t_{i, k})\}$.

The set is ordered by ascending time-stamp values. A typical YET may comprise thousands to millions of trials, and each trial may have approximately between 800 to 1500 `event time-stamp' pairs, based on a global event catalogue covering multiple perils. The representation of the YET is shown in Equation \ref{equation1}, where $i = 1, 2, \dots$ and $k = 1, 2, \dots, 1500$. 
\begin{equation}
\label{equation1}
YET	= \{ T_i = \{(E_{i, 1}, t_{i, 1}), \dots, (E_{i, k}, t_{i, k})\} \}
\end{equation}

\subsubsection*{Event Loss Tables}
This is a representation of collections of specific events and their corresponding losses with respect to an exposure set denoted as $ELT$. Each record in an ELT is denoted as $EL_{i} = \{E_{i}, l_{i}\}$ and the financial terms associated with the ELT are represented as a tuple $\mathcal{I} = (\mathcal{I}_{1}, \mathcal{I}_{2}, \dots)$. 

A typical aggregate analysis may comprise 10,000 ELTs, each containing 10,000-30,000 event losses with exceptions even up to 2,000,000 event losses. The ELTs can be represented as shown in Equation \ref{equation2}, where $i = 1, 2, \dots , 30,000$.
\begin{equation}
\label{equation2}
ELT=\left\{
	\begin{array}{l c l}
	EL_{i}			&	=	&	\{E_{i}, l_{i}\},\\
	\mathcal{I} 		&	=	&	(\mathcal{I}_{1}, \mathcal{I}_{2}, \dots)
	\end{array}\right\}
\end{equation}

\subsubsection*{Portfolio}
A Portfolio, denoted as $PF$ contains a group of Programs, denoted as $P$ represented as
$
PF = \{P_{1}, P_{2}, \cdots, P_{n}\}
$
with $n = 1, 2, \dots, 10$.

Each Program in turn covers a set of Layers, denoted as $L$, cover a collection of ELTs under a set of layer terms. A single layer $L_i$ is composed of two attributes. Firstly, the set of ELTs 
$
\mathcal{E} = \{ELT_1, ELT_2, \dots, ELT_j\}, 
$
and secondly, the Layer Terms, denoted as 
$
\mathcal{T} = (\mathcal{T}_{OccR}, \mathcal{T}_{OccL}, \mathcal{T}_{AggR}, \mathcal{T}_{AggL}).
$

A typical Layer covers approximately 3 to 30 individual ELTs and is represented as shown in Equation \ref{equation3}, where $j = 1, 2, \dots, 30$.
\begin{equation}
\label{equation3}
L=\left\{
	\begin{array}{l c l}
	\mathcal{E}	& =	& \{ELT_1, ELT_2, \dots, ELT_j\}, \\
	\mathcal{T}	& = & (\mathcal{T}_{OccR}, \mathcal{T}_{OccL}, \mathcal{T}_{AggR}, \mathcal{T}_{AggL})
	\end{array}\right\}
\end{equation}

Given the above three inputs, Aggregate Risk Analysis is shown in Algorithm \ref{algorithm1}. The data tables, $YET$, $ELT$ and $PF$, are loaded into memory.The analysis is performed in four steps for each Layer and for each Trial in the YET and a Year Loss Table ($YLT$) is produced. 

\begin{algorithm} 
\caption{Aggregate Risk Analysis}
\label{algorithm1}
\SetAlgoLined
\DontPrintSemicolon

\SetKwInOut{Input}{Input}
\SetKwInOut{Output}{Output}

\BlankLine

\Input{$YET$, $ELT$, $PF$}
\Output{$YLT$}

\BlankLine

\For{each Program, $P$, in $PF$}{
	\For{each Layer, $L$, in $P$}{
		\For{each Trial, $T$, in $YET$}{
			\For{each Event, $E$, in $T$}{
				\For{each $ELT$ covered by $L$}{
					Lookup $E$ in the $ELT$ and find corresponding loss, $l_{E}$\;
					Apply Financial Terms to $l_{E}$\;
					$l_{T} \leftarrow$ $l_{T}$ + $l_{E}$\;
				}
				Apply Occurrence Financial Terms to $l_{T}$\;	
				Apply Aggregate Financial Terms to $l_{T}$\;
			}
		}
	}
}			
Populate $YLT$ using $l_{T}$\;
\BlankLine
\end{algorithm}

In the first step, each event of a trial and its corresponding event loss in the set of ELTs associated with the Layer is determined (line 6). In the second step, a set of contractual financial terms are applied to each loss value of the Event-Loss pair extracted from an ELT to the benefit of the layer (line 7). For this the losses for a specific event's net of financial terms $\mathcal{I}$ are accumulated across all ELTs into a single event loss (line 8). In the third step, the event loss for each event occurrence in the trial, combined across all ELTs associated with the layer, is subject to occurrence terms (line 10). In the fourth step, aggregate terms are applied (line 11). 

The financial terms applied on the loss values combined across all ELTs associated with the layer are Occurrence and Aggregate terms. Two occurrence terms, namely (i) Occurrence Retention, denoted as $\mathcal{T}_{OccR}$, which is the retention or deductible of the insured for an individual occurrence loss, and (ii) Occurrence Limit, denoted as $\mathcal{T}_{OccL}$, which is the limit or coverage the insurer will pay for occurrence losses in excess of the retention are applied. Occurrence terms are applicable to individual event occurrences independent of any other occurrences in the trial. The occurrence terms capture specific contractual properties of 'eXcess of Loss' \cite{excessofloss-1} treaties as they apply to individual event occurrences only. The event losses net of occurrence terms are then accumulated into a single aggregate loss for the given trial. The occurrence terms are applied as $l_{T} = min ( max ( l_{T} - \mathcal{T}_{OccR} ), \mathcal{T}_{OccL})$.

Two aggregate terms, namely (i) Aggregate Retention, denoted as $\mathcal{T}_{AggR}$, which is the retention or deductible of the insured for an annual cumulative loss, and (ii) Aggregate Limit, denoted as $\mathcal{T}_{AggL}$, which is the limit or coverage the insurer will pay for annual cumulative losses in excess of the aggregate retention are applied. Aggregate terms are applied to the trial's aggregate loss for a layer. Unlike occurrence terms, aggregate terms are applied to the cumulative sum of occurrence losses within a trial and thus the result depends on the sequence of prior events in the trial. This behaviour captures contractual properties as they apply to multiple event occurrences. The aggregate loss net of the aggregate terms is referred to as the trial loss or the year loss. The aggregate terms are applied as $l_{T} = min ( max ( l_{T} - \mathcal{T}_{AggR} ), \mathcal{T}_{AggL})$.
 
The output of the analysis is a loss value associated with each trial of the YET. A reinsurer can derive important portfolio risk metrics such as the Probable Maximum Loss (PML) \cite{pml1} and the Tail Value-at-Risk (TVaR) \cite{tvar1} which are used for both internal risk management and reporting to regulators and rating agencies. Furthermore, these metrics flow into a final stage of the risk analytics pipeline, namely Enterprise Risk Management, where liability, asset, and other forms of risks are combined and correlated to generate an enterprise wide view of risk.
 
Additional functions can be used to generate reports that will aid actuaries and decision makers. For example, reports presenting Return Period Losses (RPL) by Line of Business (LOB), Class of Business (COB) or Type of Participation (TOP), Region/Peril losses, Multi-Marginal Analysis and Stochastic Exceedance Probability (STEP) Analysis.

\section{Experimental Studies}
\label{results}
In this section the experimental platform, implementation and the results obtained are presented. 

The experimental platform employed in this research comprises 1027GR-TRF Supermicro nodes. Each node contains two Intel Xeon E5-2620 v2 processors, each with six cores, operating at 2.1 GHz and 32 GB of DDR3 SDRAM memory at 1600 MHz. Each node has a Mellanox ConnectX-3 VPI single-port InfiniBand adapter (InfiniBand FDR) as well as a Mellanox ConnectX-2 VPI single-port adapter (InfiniBand QDR). The nodes are connected either by a Mellanox switch MTS3600 with QDR compatibility (a maximum rate of 40Gb/s) or by a Mellanox Switch SX6025, which is compatible with InfiniBand FDR (a maximum rate of 56Gb/s). One NVIDIA Tesla K20 GPUs is available for acceleration on each node. Additionally, one Supermicro server with identical processors was populated with 4 NVIDIA Tesla K20 GPUs for the purpose of comparison. The CentOS 6.4 operating system is used, and the Mellanox OFED 2.1-1.0.0 (InfiniBand drivers and administrative tools) was used at the servers along with CUDA 6.0. 

The financial risk case study was implemented as follows. A single thread was employed for the computations of each trial of the application. ELTs corresponding to a Layer were implemented as direct access tables to facilitate fast lookup of losses corresponding to events. In our implementation, each ELT is considered as an independent table; therefore, in a read cycle, each thread independently looks up its events from the ELTs. All threads within a block access the same ELT. The device global memory stores all data required for the analysis. Chunking, which refers to processing a block of events of fixed size (or chunk size) for the efficient use of shared memory is employed to optimise the implementation. The computations related to the events in a trial and financial terms are chunked. The financial terms are stored in the streaming multi-processor's constant memory. In this case, chunking reduces the number of global memory update and global read operations. 

\begin{figure}[t]
\centering
	\subfloat[Execution time on a single GPU using varying number of blocks per thread]{\label{figure8a}\includegraphics[width=0.49\textwidth]{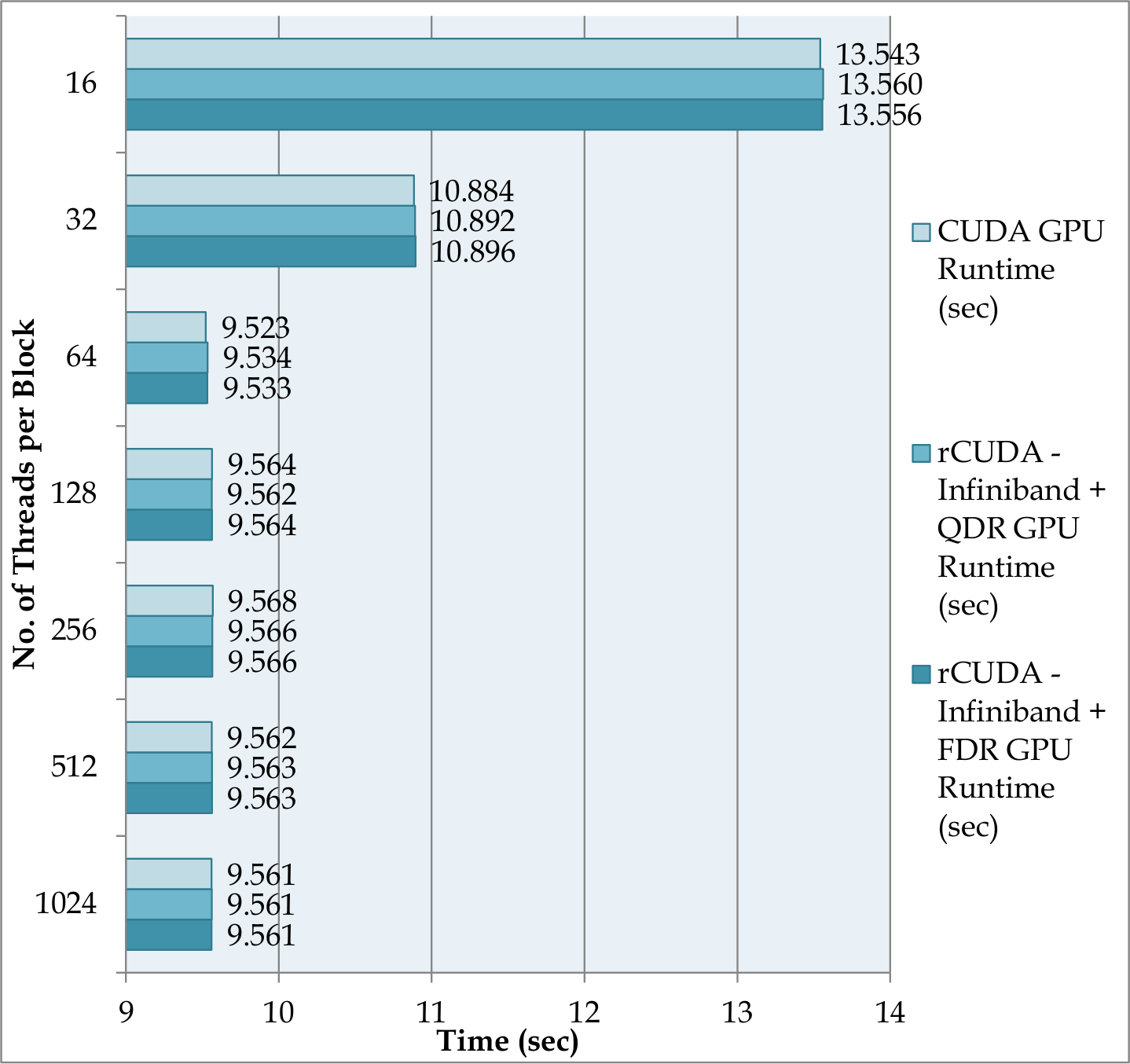}} \\
	\subfloat[Time for copying data on a single GPU using varying number of blocks per thread]{\label{figure8b}\includegraphics[width=0.49\textwidth]{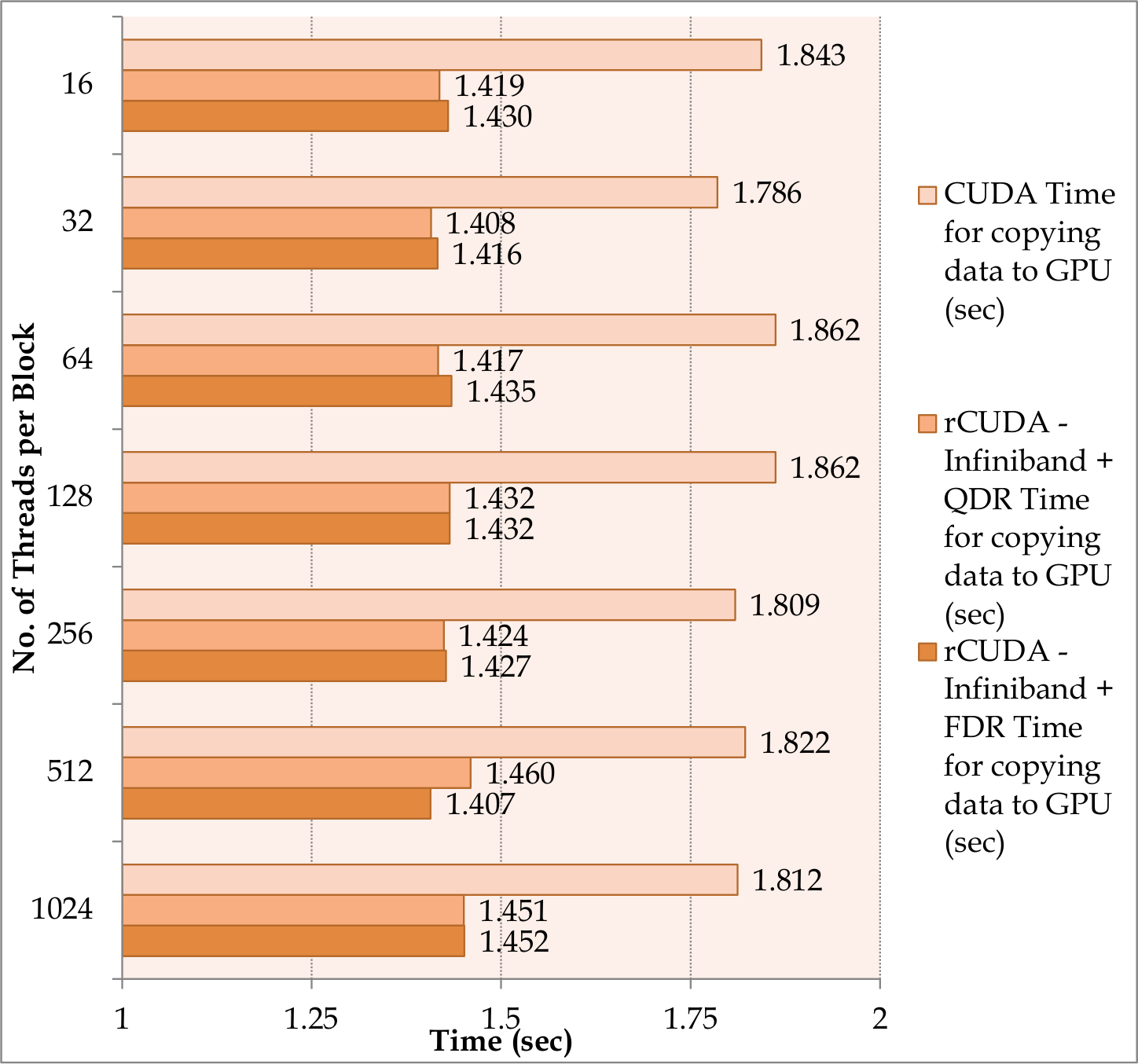}}
\caption{Experimental results from executing the financial risk application on a single GPU}
\label{figure8}
\end{figure}

Aggregate risk analysis was executed on industry size input to obtain the results presented in this paper. The YET comprising one million trials, with each trial simulating 1,000 events on an average, and one layer with 16 ELTs was employed. The results compare the time taken by CUDA on physical (local) GPUs and by rCUDA on virtualised (remote) GPUs using InfiniBand QDR and FDR. All timing results are an average of five executions. We noted that when using rCUDA the standard deviation was less than 0.002 seconds for the execution time and less than 0.004 seconds for data transfer time. Such a low standard deviation highlights the robustness of the rCUDA framework in providing consistent results on the experimental platform. 

Figure \ref{figure8} shows the execution of aggregate risk analysis on a single GPU (physical if CUDA or virtualised if rCUDA). As expected, the execution time decreases as the number of blocks per thread are increased (refer Figure \ref{figure8a}). It is interesting to note that the performance of virtualised GPUs using rCUDA is very close to that offered by the physical GPU. When 16 to 64 threads per block are employed, there is only a 0.1\% overhead, which is negligible, when using rCUDA. Beyond 64 threads the rCUDA execution time is very close to the CUDA time. The data transfer time from the CPU to the GPU is shown in Figure \ref{figure8b}. Surprisingly, the time taken to copy using rCUDA is faster by 20\% over CUDA. The reason is that rCUDA achieves a higher bandwidth than CUDA for memory copies from host memory to GPU memory when pageable memory is leveraged. This is due to the use of internal pinned memory
buffers between the different stages of the rCUDA pipeline~\cite{rcuda-new}.

\begin{figure}[t]
\centering
	\subfloat[Execution time on four GPUs using varying number of threads per block]{\label{figure8c}\includegraphics[width=0.49\textwidth]{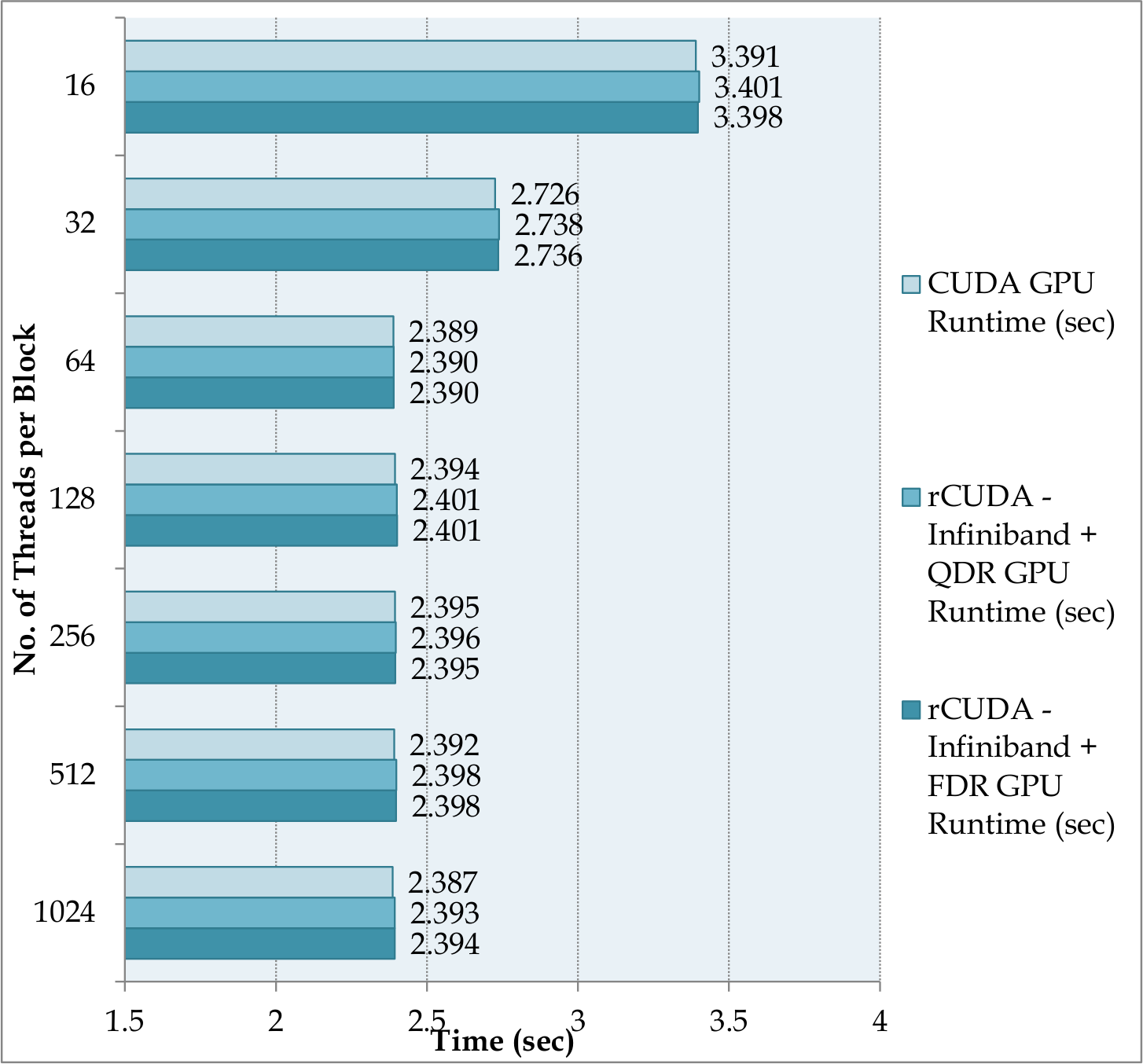}} \\ 
	\subfloat[Time for copying data on four GPUs using varying number of threads per block]{\label{figure8d}\includegraphics[width=0.49\textwidth]{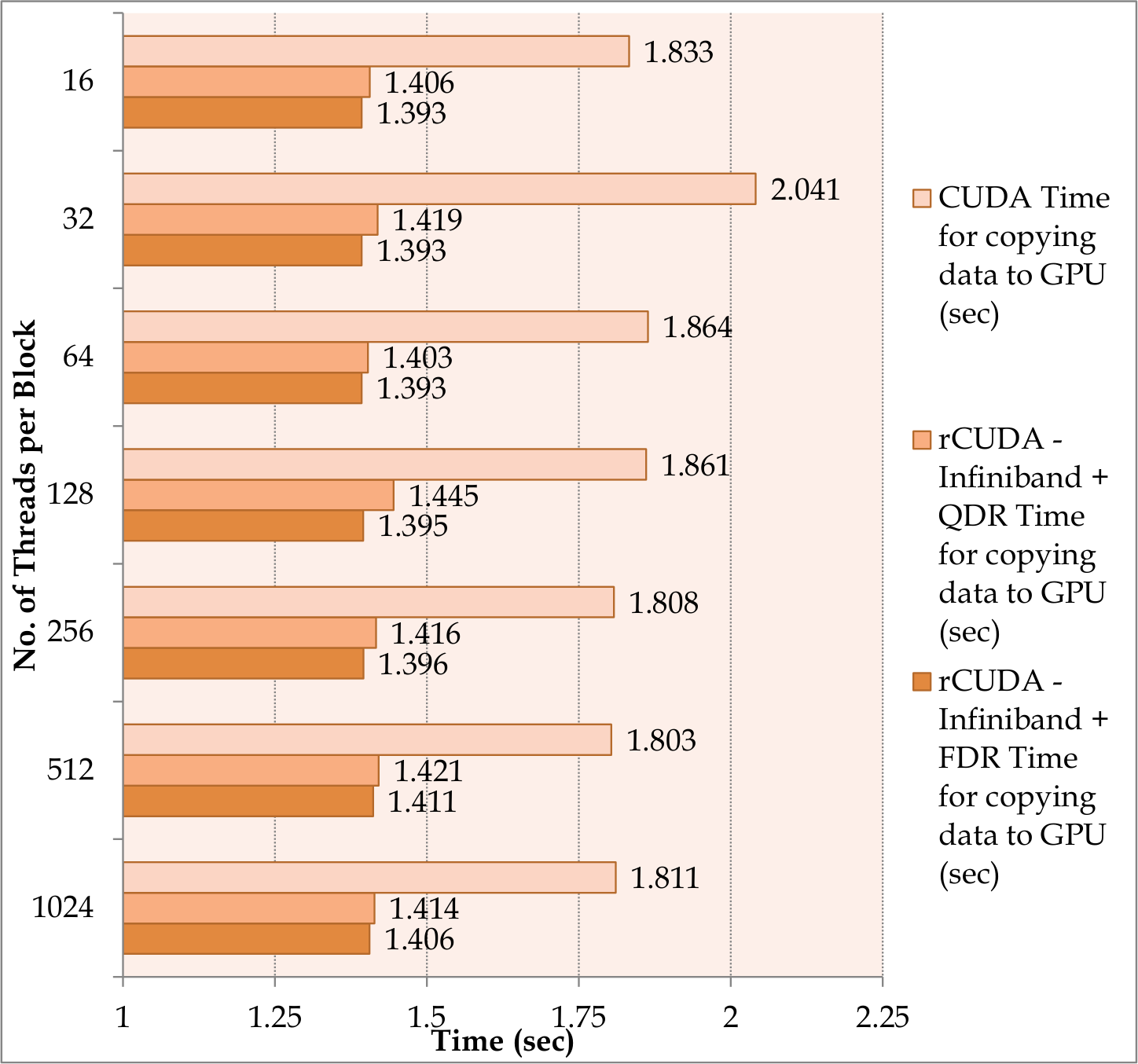}}
\caption{Experimental results from executing the financial risk application on four GPUs}
\label{figure9}
\end{figure}

To further validate the feasibility of rCUDA on multiple GPUs,  aggregate risk analysis was executed on four GPUs. In the case of CUDA, the node owning four GPUs was used. In the case of rCUDA, four different GPU servers owning one GPU each were leveraged. Again, the results shown in Figure \ref{figure9} bear resemblance to the results from a single GPU. The performance of rCUDA is very close to that of CUDA. There is only a 0.2\% overhead introduced by rCUDA when 16 to 32 threads per block are employed, beyond which there are no overheads (refer Figure \ref{figure8c}). The data transfer time from the CPUs to the GPUs is shown in Figure \ref{figure8d}.    

\begin{figure}[t]
\centering
	\subfloat[Execution time on four GPUs]{\label{figure8e}\includegraphics[width=0.49\textwidth]{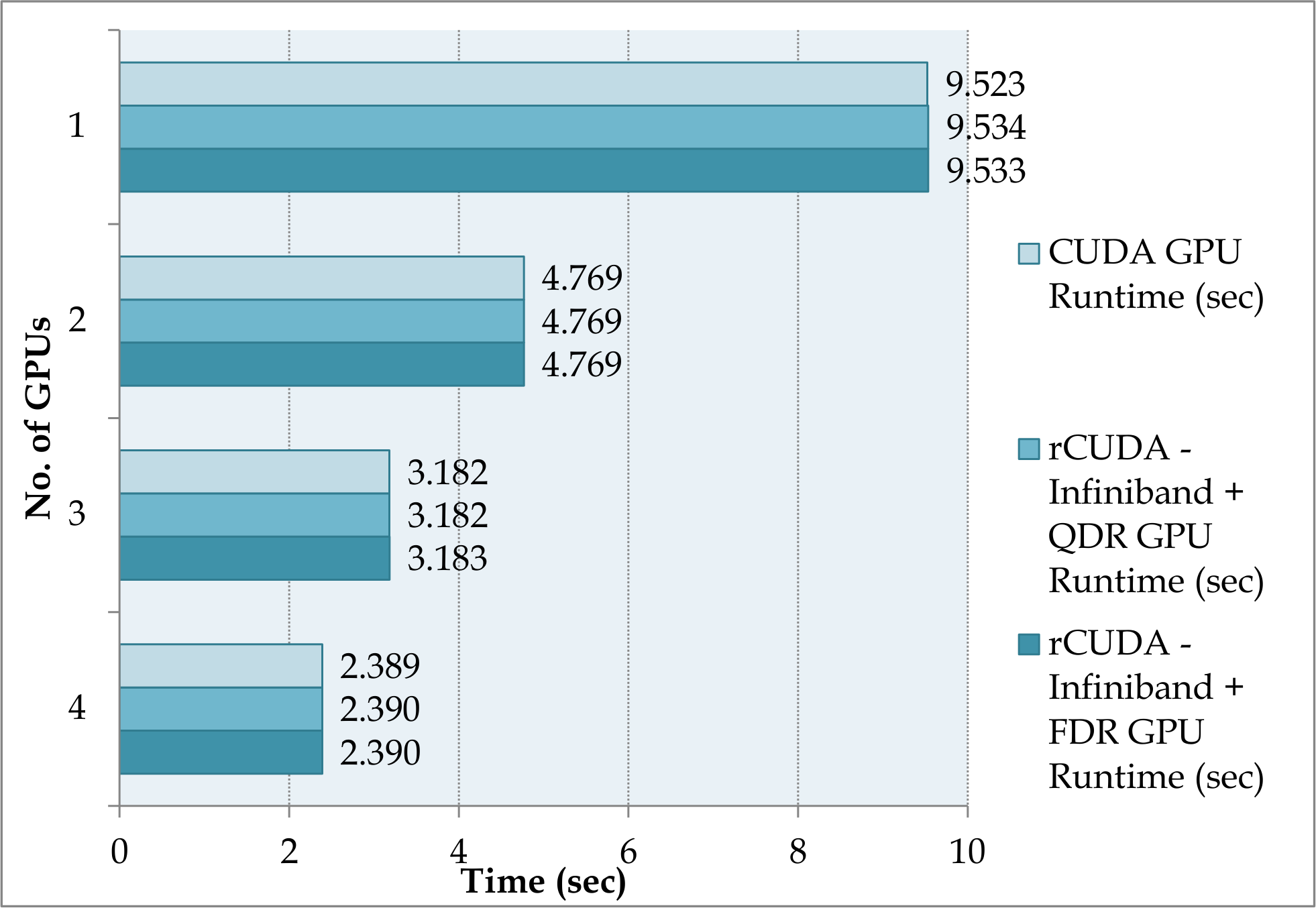}}\hspace{10pt}
	\subfloat[Execution time on varying number of virtualised GPUs ]{\label{figure8f}\includegraphics[width=0.49\textwidth]{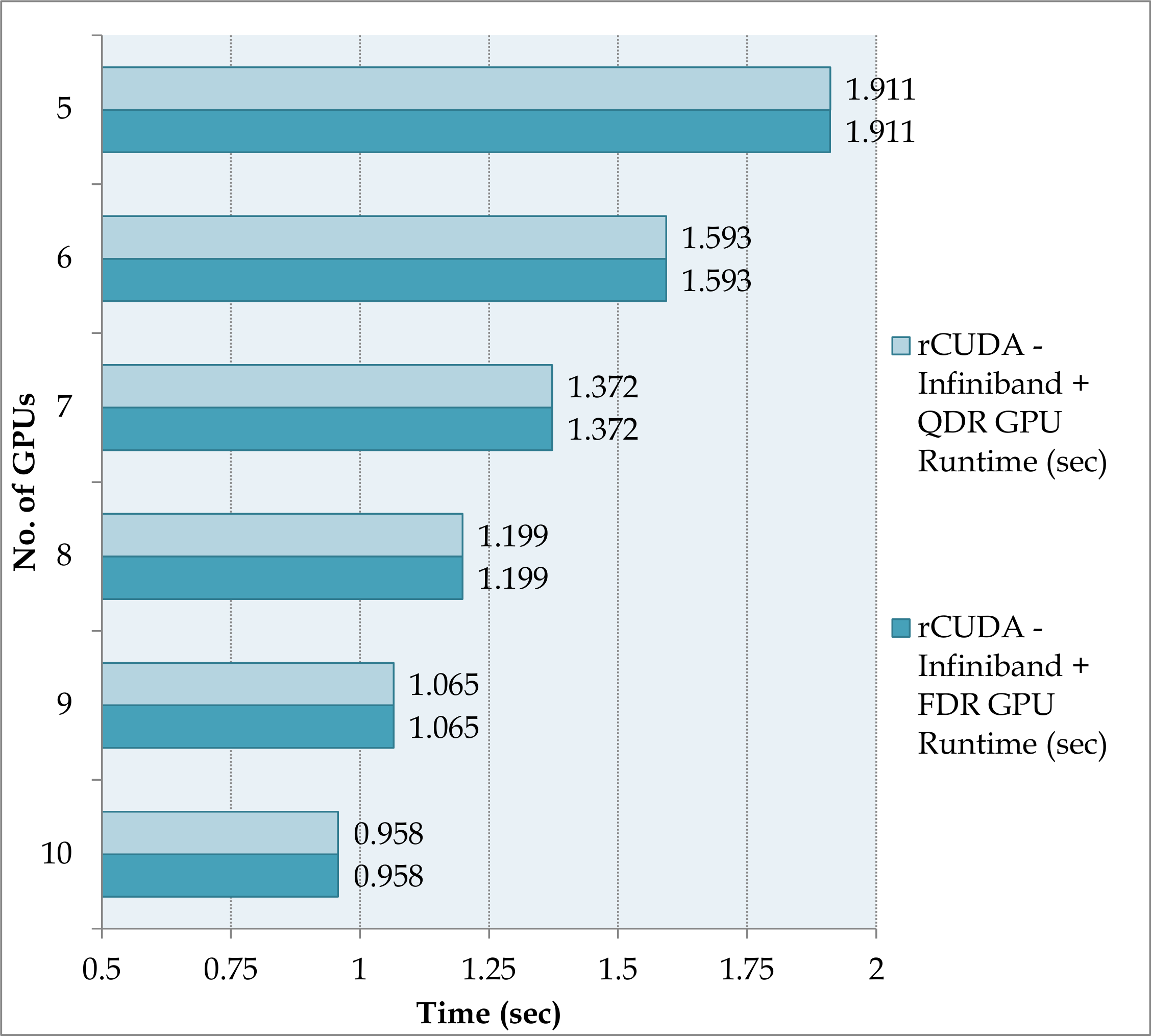}} \\
\caption{Experimental results from executing the financial risk application on 64 threads per block}
\label{figure10}
\end{figure}

Figure \ref{figure8e} shows the timing results using 64 blocks per thread when CUDA and rCUDA are employed. Similar to results shown in Figure \ref{figure9}, CUDA executions were carried out in the node with four GPUs whereas rCUDA executions used four different servers with one GPU each. Figure \ref{figure8e} shows that the overhead of using rCUDA is negligible when compared to local CUDA. However, the results in Figure~\ref{figure8f} are interesting. Using rCUDA it is possible to provide a single application with more GPUs than those available in a single node. Here we note that with rCUDA it is possible to assign a single application all the GPUs available in the cluster. The financial risk analysis application is assigned up to ten GPUs in ten different remote servers. It is evident that the performance of the application continues to improve with additional GPUs. In this case, the application time is brought down to less than 1 second, making it possible for use in real-time settings, such as on-line pricing.

To summarise the results, the experimental studies on aggregate risk analysis highlight that: (i) rCUDA achieves near to similar performance as that of CUDA. In our experiments, for 64 threads per block on a single GPU CUDA took 9.523 seconds, where as rCUDA required 9.533 seconds. The execution time indicates that for an industry size simulation of the risk problem, there was an overhead of only 0.01 seconds in rCUDA. (ii) Consistent results can be obtained from rCUDA. The standard deviation of multiple executions was negligible. (iii)  rCUDA allows for the use of as many virtual GPUs (as real GPUs) are available in the cluster. This not only facilitates boosting the performance of applications, but also the larger number of GPUs that can be accessed over traditional CUDA.

\section{Related Work}
\label{relatedwork}
High Performance Computing (HPC) solutions are exploited in the financial risk industry to accelerate the underlying computations of applications. This reduces overall execution times making such applications fit for real-time use. Solutions range from small scale clusters \cite{smallcluster1, smallcluster2} to large supercomputers \cite{supercomputer1, supercomputer2}. More recently, hardware accelerators with multi-core and many-core processors are employed. For example, financial risk applications are accelerated on Cell BE processors \cite{cellbe-1, cellbe-2}, FPGAs \cite{fpga3, fpga4} and GPUs \cite{gpu3, gpu5}. 

HPC clusters offering heterogeneous solutions by using hardware accelerators, such as GPUs, along with processors on nodes are feasible \cite{hetcluster-1, hetcluster-2}. One way to set up such clusters would be to incorporate a GPU on each node of the cluster. However, this set up will not be efficient because of (i) relatively high performance/cost ratio of GPUs, (ii) higher power consumption of a node hosting GPUs \cite{gpu-power1}, and (iii) under utilisation of GPUs when available on all nodes of the cluster.

An alternate set up to the one considered above is to employ fewer GPUs than the number of cluster nodes and efficiently share them between the nodes. Such a solution will be appealing to the industry to reduce costs due to investing in a lot of hardware and consequentially in maintenance, space, cooling and power consumption. But this solution poses the challenge of efficiently providing Acceleration-as-a-Service (AaaS) to the nodes of a cluster when it is requested for without significant loss of performance. Virtualisation of GPUs \cite{virt-1,virt-2} lies at the heart of the solution and efficiently using it can surmount the challenge. 

A number of frameworks, such as rCUDA \cite{rcuda-new,rcuda-infiniband}, V-GPU\footnote{https://github.com/zillians/platform\_manifest\_vgpu}, GridCuda \cite{gridcuda}, DS-CUDA \cite{dscuda}, and Shadowfax II~\cite{Shadowfax} are available for GPU virtualisation. rCUDA features CUDA 6.5 and provides specific communication support for TCP/IP compatible networks as well as for InfiniBand fabrics. V-GPU supports CUDA 4.0 and there are no publicly available versions that can be tested. GridCuda supports CUDA 2.3, an old version, and again is not publicly available for testing. DS-CUDA supports CUDA 4.1 and includes specific communication support for InfiniBand. However, DS-CUDA is limited in that it does not support data transfers with pinned memory. A stable version of Shadowfax II is not available and is still under development. The information available publicly does not reflect the current development status.

The rCUDA framework reported in this paper facilitates the set up in which fewer GPUs need to be used than the number of nodes in a cluster environment. AaaS is achieved such that acceleration can be requested as a service by any node in the cluster and concurrent access of a GPU by multiple cluster nodes is possible.

\section{Conclusions}
\label{conclusions}
An application can be easily accelerated in a cluster or a supercomputer set up that has a hardware accelerator, such as a GPU, on each of its nodes. However, this is not pragmatic since the GPUs will be under utilised in addition to the high performance/cost ratio and power consumption. An alternate set up is to use a fewer number of GPUs such that acceleration can be requested on demand by a node. To facilitate this, a framework that can efficiently handle the virtualisation of GPUs is required.

In this paper, we employed the rCUDA framework as an approach to obtain \textit{Acceleration-as-a-Service (AaaS)} in a cluster. Each node can request acceleration when required and a virtual GPU is made available in response to the request. The framework ensures that the physical GPU is not locked exclusively to the node requesting acceleration but can be shared through multiple virtual GPUs. The feasibility of the framework was tested on a real-world application used in the financial risk industry. The results are encouraging and rCUDA achieves similar performance in comparison with CUDA and an application can draw from a large pool of GPUs to boost performance.

\section*{Acknowledgement}
This research was supported by an NVIDIA award, the EPSRC EP/K015745/1 grant, and the Generalitat Valenciana PROMETEOII/2013/009 grant of the
PROMETEO program phase II. The authors acknowledge the generous support of Mellanox Technologies.


\end{document}